\documentclass[]{aa}
\usepackage{graphicx}
\usepackage{txfonts}
\usepackage{longtable}
\begin{document}
   \title{Formation of water and methanol in star forming molecular clouds}
   \author{A. Das\inst{1}, K. Acharyya \inst{2}, S. Chakrabarti \inst{3,1}, 
and S. K. Chakrabarti \inst{2,1} }
\offprints{Prof. Sandip K. Chakrabarti}
\institute{Indian Centre for Space Physics, Chalantika 43, Garia Station Rd., Kolkata, 700084, India\\
   \email{ankan@csp.res.in}
\and
S. N. Bose National Centre for Basic Sciences, Salt Lake,
              Kolkata 700098, India\\
             \email{acharyya@bose.res.in, chakraba@bose.res.in}
\and
Maharaja Manindra Chandra College, 20 Ramakanta Bose lane, Kolkata 700003, India\\
   \email{sonali@csp.res.in}}
                                                                                                                

   \date{Received ; accepted }
\abstract
   {} 
{We study the formation of water and methanol in the dense cloud conditions to 
find the dependence of its production rate on the binding energies, reaction 
mechanisms, temperatures, and grain site number. We
wish to find the effective grain surface area available for chemical reaction
and the effective recombination timescales as functions of grain and gas parameters.
}
{We used a Monte Carlo simulation to follow the chemical processes 
occurring on the grain 
surface. We carried out the simulations on the Olivine grains of different 
sizes, 
temperatures, gas phase abundances and different reaction mechanisms. 
We consider H, O, and CO as the accreting species from the gas phase and 
allow ten chemical reactions among them on the grains.}
{We find that the formation rate of various molecules is strongly dependent on the 
binding energies. When the binding energies are high, it is very difficult 
to produce significant amounts of the molecular species. Instead, the grain is found to be
full of atomic species. The production rates are found to depend on the number
density in the gas phase. When the density is high, the production of 
various molecules on the grains is small as grain sites are quickly filled up by 
atomic species. If both the Eley-Rideal and Langmuir-Hinselwood 
mechanisms are considered, then the production rates are at this maximum and the grains are filled up relatively 
faster. Thus, if allowed, the Eley-Rideal mechanism can also play a major role and more so when 
the grain is full of immobile species. We show that the concept of the effective grain surface area, 
which we introduced in our earlier work, plays a significant role in grain chemistry.}
{We compute the abundance of water and methanol and show that the results strongly depend on the
density and composition in the gas phase, as well as various grain parameters.  
In the rate equation, it is generally assumed that the recombination 
efficiencies are independent of the grain parameters, and the surface coverage. 
Presently, our computed parameter $\alpha$ for each product is found to depend on 
the accretion rate, the grain parameters and the surface coverage of the grain. 
We compare our results obtained from the rate equation and the one from
the effective rate equation, which includes $\alpha$. A comparison  of our results
with the observed abundance shows very good agreement.}

\titlerunning{Water and methanol formation}
\authorrunning{Das, Acharyya, Chakrabarti and Chakrabarti}
\maketitle

\section{Introduction}

Formation of complex organic molecules in the interstellar medium (ISM) is an active 
subject of research. Out of a host of organic and inorganic molecules that 
have been
observed, water and methanol are most certainly two very important organic species found 
both in the gas and solid phases of ISM. The abundance of these species in the various regions 
of ISM is also different. It was thought for a long time that the water is one of the possible
reservoirs of elemental oxygen in the gas phase, but recent Submillimeter Wave Astronomy Satellite 
(SWAS) observation found surprisingly a low abundance of water. 
Snell {\it {et al.}} (2000) find that the abundance of water relative to H$_2$ in Orion 
and M17 cloud is between $10^{-10}$ to $8 \times 10^{-10}$. The water abundance in the hot cores 
range from  $10^{-6}$ to $10^{-4}$ (van Dishoeck \& Helmich 1996, Helmich {\it {et al.}} 1996;
Boogert \& Ehrenfreund, 2004). The abundance of water  on grains with respect to the total H column
density is typically $10^{-4}$ and is the most abundant component (Tielens et al. 1991).
Similarly, interstellar methanol has three types of abundance profile: flat profiles at CH$_3$OH/H$_2$ 
$\sim 10^{-9}$ for the coldest sources, profiles with a jump in its abundance from $\sim 10^{-9}$ 
to $\sim 10^{-7}$ for the warmer sources, and flat profiles at a $\sim$ few $10^{-8}$ for the 
hot cores (van der Tak {\it {et al.}} 2000). On the grain surface, the methanol abundance 
varies from  $5 \%$ to $30 \%$ with respect to H$_2$O. In some sources, such as SgrA and Elias 16,
the abundance is even less (Gibb {\it at al.}, 2000). 
The observed abundance for methanol along the line of sight towards high mass and low-mass proto-stars 
is between $0.2 - 2 \times 10^{-5}$ (Gibb et al. 2004; Pontoppidan et al. 2003, 2004).
Much higher methanol abundances are found to be associated with the 
outflows in the regions of low mass star formation, L1157-MM and NGC1333-IRAS2 
$2 \times 10^{-5}$ and $2 \times 10^{-6}$ (Bachiller \& Perez Gutierrez, 1997; Bachiller
et al. 1998). In other words, either on the grain surfaces or in the 
hotter region, abundances of theses species are high. This correlation suggests that
these species perhaps originate from grains and their productions in the gas phase are
inadequate. Therefore, the understanding of the formation of water and methanol on grain surfaces 
is of primary importance. 

The grain surface reactions were first introduced to explain the formation of 
molecular hydrogen (Hollenbach \& Salpeter, 1970). Since then it has been used very extensively by several 
authors (Watson \& Salpeter 1972ab; Allen \& Robinson 1975, 1976, 1977; Tielens \& Hagen 1982; 
Hasegawa \& Herbst 1992; Charnley 2001; Stantcheva {\it et al.} 2002, 
Green {\it et al. 2001}; Biham {\it et al.} 2001; Stantcheva {\it et al.} 2002). 
These studies mainly belong to two different categories, the deterministic
approach and the stochastic approach. 
In the deterministic approach, one can completely determine the time evolution of the system once the 
initial conditions are known. The rate equation method belongs to this category. This method is very 
extensively used by several authors to study the grain surface chemistry 
(Hasegawa \& Herbst 1992; Roberts {\it et al.} 2002; Acharyya et al. 2005). However, this method is 
only applicable when there are large numbers of reactants on the grain surface. Given that 
the interstellar medium is very dilute, very often this criteria is not fulfilled and this method cannot 
be applied. But this method is computationally faster and can very easily be coupled with the gas phase 
reactions. In the stochastic approach, fluctuations in the surface abundance due to the 
statistical nature of the grain is preserved. The Monte Carlo method and the Master equation methods 
belong to this category. Both these methods are used by several authors (Charnley 2001; 
Stantcheva {\it et al.} 2002, Green {\it et al. 2001}; Biham {\it et al.} 2001; Stantcheva {\it et al.} 2002 ).
Its major disadvantage is that it takes enormous computational time. Coupling of the Monte Carlo method 
(for grain surface reactions) and rate equation method (for gas phase reactions) is extremely difficult.
Although the Master equation method can be coupled with the rate equations, but it is disadvantageous because
one has to solve a large number (ideally infinite) of reactions.

\begin {figure}
\vskip 0.6cm
\centering{
\includegraphics[width=6cm]{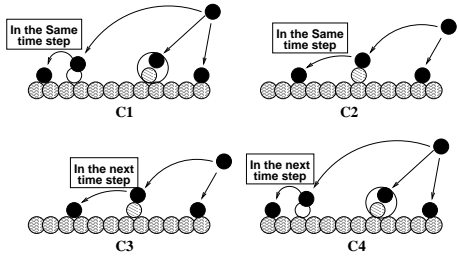}}
\caption{Cartoon diagram to show the different reaction schemes (C1-C4) that are considered in our calculation. 
Black and shaded circles are representatives of reactive species and the clear 
circles are for nonreactive species. A circle around two such species indicates that the new species is
created due to Eley-Rideal scheme.}
\label{fig-1}
\end {figure}
Recently, Chang {\it et al. } (2005) have argued that the stochastic methods used so far can also lead to error 
because the rate of reaction is determined by the rate of hopping (or tunneling) of a hydrogen atom from 
one site to the nearest neighboring site multiplied by the probability of finding a reactant partner in this 
site. This is also an average treatment since, on any given grain, the reactant partner is unlikely to lie 
in the nearest-neighboring site. They used a continuous random work technique to study the formation of 
molecular hydrogen. Chakrabarti {\it et al. (2006ab)} used a similar method that keeps track of each 
individual reactant and their movements and calculated the effective grain surface area involved in the 
formation of molecular hydrogen in the interstellar clouds. Chakrabarti {\it {et al.}} (2006, 
hereafter referred as Paper-I) illustrated that the formation rate per unit grain site 
itself strongly depends on the nature and size (i.e., no. of sites) of the grains for a given 
gas phase condition (abundance, temperature, etc.). In Paper-I, this was demonstrated only 
for H forming H$_2$ molecules.  In the present paper, we carry out a similar analysis 
where we consider the accretion of H, O, and CO onto the grain surface and show that the 
formation rates of water and methanol are indeed dependent on the intrinsic 
and extrinsic parameters of the system. The plan of this paper is the following. In the next 
section, we discuss various mechanisms of reactions on the surface. In Section 3, we 
discuss the procedure of our computation. In Section 4, we describe various models 
used for our simulations. In Section 5, we present our results. Finally, in Section 6, 
we draw our conclusions.
\begin {figure}
\vskip 0.6cm
\centering{
\includegraphics[width=7cm]{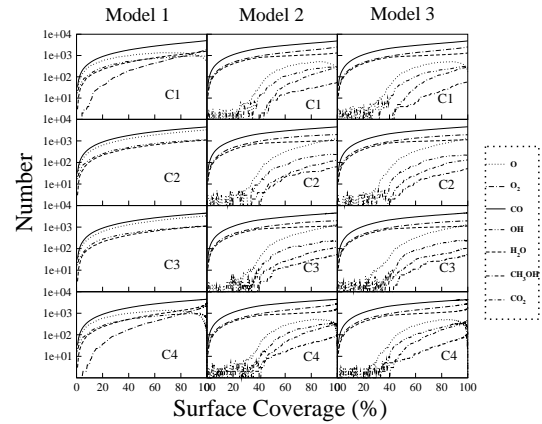}}
\caption{Evolutions of the number of a few selected species on a 
grain having $10^4$ sites are shown with respect to the surface 
coverage of a mono-layer (along X-axis), for various sets of binding energies  (Models 1-3)
and for different mechanisms (C1-C4). These simulations were carried out for high abundances 
of the accreting species (Table 1). Different styles of the curves are marked with the species names on the right.
Note that methanol and CO$_2$ are absent in the Model 1 simulation where the binding energy was very high.}
\label{fig-2}
\end {figure}
 
\section{Mechanisms of reactions on grain surfaces}

A thorough understanding of the surface reaction mechanisms requires 
knowing of the basic physical processes involved when the gas phase 
atoms and molecules interact with the grains. The first step is accretion, i.e., 
landing various species onto a grain surface. In our case, only H, O, and CO 
are taken as the accreting species onto the grain surface. In the next step, 
the accreted species will react to form various new species. There are two reaction 
schemes, the Langmuir-Hinselwood (LH) mechanism and the Eley-Rideal (ER) mechanism. 
In the LH scheme, the gas phase species accretes onto a grain and becomes equilibrated 
with the surface before it reacts with another atom or molecule, and in the ER 
reaction scheme, the incident gas phase species collides directly with an adsorbed
species on the surface and reacts with that species. In such a mechanism, 
the reactant generally does not become trapped at the surface and it is unlikely to be sensitive 
to the surface temperature (Farebrother et al. 2000). In our study, we considered 
that the reaction on a surface can occur through both LH and ER mechanisms. However, we assume
that the molecules are trapped after reaction due to their high binding energy. 

The binding energy of the incoming species strongly depends on the species itself and on the way the interaction 
proceeds. The incoming species might get trapped in a shallow potential well in a physisorbed site. The interaction
is mainly due to mutually induced dipole moments or it might also form a strong covalent bond. Recent studies have 
found evidence of both physisorption and chemisorption processes taking place on a grain surface. However, 
for the chemisorption, a higher kinetic energy is involved, so this type of interaction is not relevant 
except in very special astrophysical conditions. We have considered only weakly bound species, i.e., physisorbed 
atoms and molecules. The typical energy for physisorption is around $0.1$ eV or $ \sim \ 1000$ K.
If $E_d$ denotes the binding energy for physical adsorption and $E_b$ the potential energy 
barrier, then $E_d$ must be greater than $E_b$ for the species to diffuse from one site to the other.
\begin {figure}
\vskip 0.6cm
\centering{
\includegraphics[width=7cm]{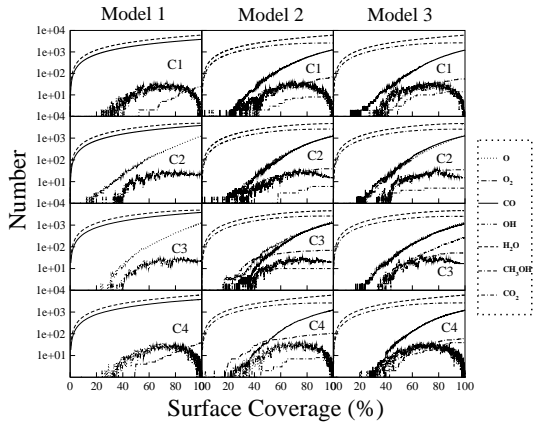}}
\caption{Same as in Fig. 2, but the low abundances of the accreting species have been chosen.}
\label{fig-3}
\end {figure}

For a chemical reaction to occur, an accreted species has to scan the grain surface in search of 
a reaction partner. There are two physical processes that can provide the mobility for 
the accreted gas phase species, the thermal hopping and tunneling. As mentioned earlier, 
Hollenbach \& Salpeter first introduced the grain surface chemistry to explain the high abundance of 
molecular hydrogen. They assumed that, within the grains, the hydrogen atoms move from site to site by a
quantum mechanical tunneling process. But from the recent experimental results of molecular hydrogen formation, 
it was found that the mobility of hydrogen on grains is primarily due to the 
thermal hopping (Pirrenello 1997ab, 1999). The tunneling time very  
dependent on the mass of the particle and the barrier thickness, a  unknown parameter.
Therefore, it is not widely used in the astrophysical models. The classical papers 
like Tielens \& Hagen (1982) and Hasegawa \& Herbst (1992) considered the tunneling for hydrogen and thermal 
hopping for other simple atoms and molecules. We considered both tunneling and thermal hopping for 
the hydrogen atoms and only thermal hopping for others. 

If we consider the time scale for hopping and tunneling for the hydrogen atom and hopping for other species,
we observe that the mobility of hydrogen atom is much greater than for the other reactive species. 
The hopping and tunneling time for hydrogen is $\sim 7.4 \times 10^{-09}$ and $2.0 \times 10^{-11}$s,
and the hopping times for atomic species like O and CO are $\sim 2.4 \times 10^{-2}$s and $ \sim 5589$s ,
respectively. Therefore, the hydrogenation reactions are the dominant reactions on grains. 
\begin{figure}
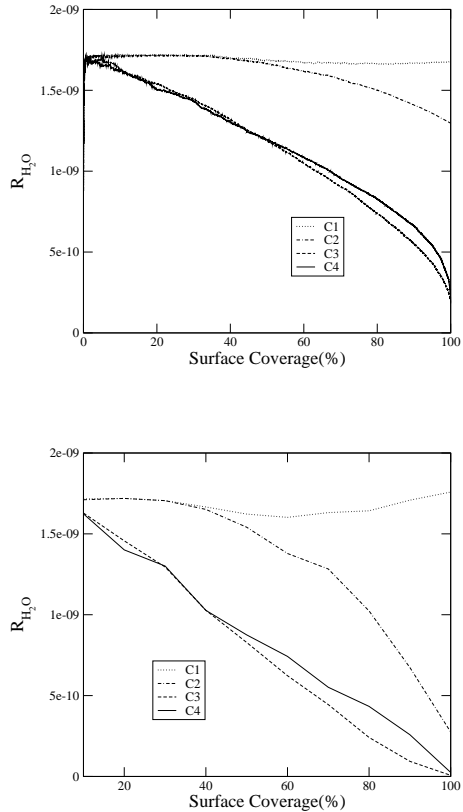

\vskip 0.6cm
\centering{
\includegraphics[width=6cm]{8422fig4.eps}
\vskip 1cm
\includegraphics[width=6cm]{8422fig5.eps}}
\caption{Rate of production of H$_2$O is shown against the surface coverage for the
four different schemes mentioned in the text. The production is high when the Eley-Rideal 
mechanism is taken into account, while the production is lower when the reaction takes 
place at the next time step. The rate is computed using (a) the total production
and total time taken at a given instant, (b) the change in the number of H$_2$O for a 
fixed change in parcentage of the surface coverage.}
\label{fig-4}
\end {figure}
\subsection {Accretion rate}
Accretion is the process by which grain receives matter.
We define the accretion rate [$r_{acc}(i)$] for a given neutral species $i$ in the units of s$^{-1}$ , as
\begin{equation}
r_{acc} (i) = s_i \sigma v_i n_{gas} (i), 
\end{equation}
where $s_i$ is the sticking coefficient (taken as $1$ for all the three species), $v_i$ the velocity 
(cm s$^{-1}$ ), $n_{gas} (i)$ is the gas phase number density (cm$^{-3}$) of the $i$-th species, 
and $\sigma$ the grain cross-section (cm$^2$). In this paper, we carry out our calculation 
usually for two different sets of gas phase number 
densities, excepts towards the end where we compared results from several number
densities. Thus, we have two sets of accretion rates. Number densities 
are taken from Stantcheva {\it {et al.}} (2002). These are listed in Table~\ref{table-1}. 

\begin{table}
\centering
\caption{Gas-phase abundances used}
\begin{tabular}{|l|l|l|}
\hline
Species & high ($cm^{-3}$)&low ($cm^{-3}$) \\
\hline
H&1.10 &1.15\\
\hline
O& 7.0&0.09\\
\hline
CO&7.5 &0.075\\
\hline
\end{tabular}
\label{table-1}
\end{table}

\subsection {Diffusion rate coefficients and binding energies}
{\begin{figure}
\vskip 0.6cm
\centering{
\includegraphics[width=6cm]{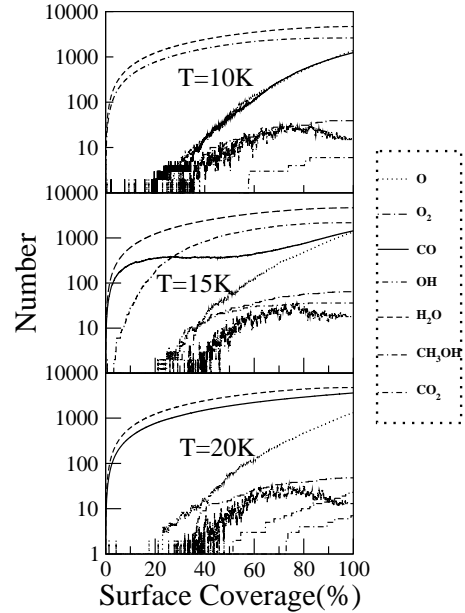}}
\caption{Rate of production of a few selected species as a function of surface coverage at different temperatures
shown for a grain having $10^4$ sites for Model 2 and C2 scheme. 
These simulations were carried out for low abundances of the accreting species. }
\end {figure}}

The diffusion rate coefficients ($k$) are the sums of the rates $t^{-1}_{diff}$ ($s^{-1}$) 
the reactive species need to traverse an entire grain ($t_{diff}$ is the diffusion time). In the present paper, we 
are interested in showing the size dependence of the recombination process. Thus, we choose a grain to contain 
$10^4$ to $ 10^5$ sites. The rate coefficient is usually multiplied by a factor $\kappa$ that 
accounts for any non zero activation barrier (Hasegawa {\it {et al.}} 1992).

The rates depend strongly on $E_b$, the barrier energy. We considered three different sets of barrier energies. 
In the first set, we used the binding energies that come from the earlier works (Allen \& Robinson, 1977; 
Tielens \& Allamandola, 1987; Hasegawa \& Herbst, 1993). We show them in Table 3. In this set, we considered that 
the hydrogen diffusion is caused by thermal hopping. Thus we ignore tunneling. In the second set, we used the 
same set of binding energies except for hydrogen diffusion procedure. Instead of hydrogen diffusion, 
we considered tunneling. Since the hydrogen diffuses much faster on grains than any other species, 
hydrogenation reaction is the dominant reaction on grains. This is why we did done 
our simulation for both thermal hopping and tunneling of hydrogen. The third set is based on recent findings 
of Pirronello {\it {et al.}} (1997, 1999) as interpreted by Katz {\it {et al.}} (1999), which show that 
atomic hydrogen moves much more slowly than what is used in various simulations. Therefore, in this set we 
used hydrogen hopping rates from Katz {\it {et al. (1999)}} and barrier energies for
other species are increased proportionately (Table~2).  This means that the ratio between
the barrier energies of these two tables for different species are exactly the same as for H 
(i.e., $2.87$ for $E_b$ and $1.066$ for $E_d$). The ratio between the desorption energies of 
Table 2 and 3 was also kept fixed for simplicity. We chose $E_b=0.3E_d$ in Table 3 for all 
species except for H.
{\begin{figure}
\vskip 0.6cm
\centering{
\includegraphics[width=6cm]{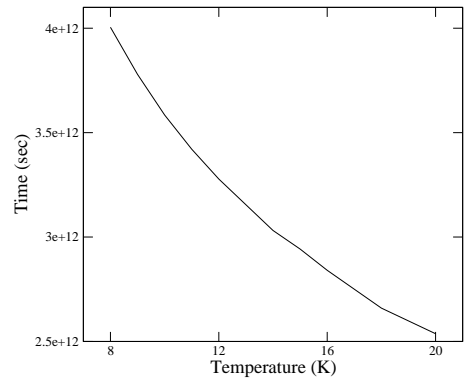}}
\caption{Plot of the time to build a mono-layer as a function of temperature
for a grain having $10^4$ sites and for Model 2 by considering the C2 scheme.
These simulations were carried out for low abundances of the accreting species.}
\end {figure}}

Using these three sets we have constructed three models. The results are shown according 
to the increasing diffusion rate of hydrogen. Thus in Model 1, we used the 
slowest diffusion rate of hydrogen and hence the binding energy of Table~2. 
In Model 2, we used the binding energy from Table 3 (below). Finally in Model 3, 
the hydrogen tunneling is considered and the binding energy is taken from Table~3 
for other species. 
\begin{table}
\centering
\caption{Energy barriers against diffusion and desorption in K for
the olivine grain (from experiment)}
\begin{tabular}{|l|l|l|}
\hline
Species &$E_b$ &$E_d$ \\
\hline
H&287&373\\
\hline
O&689&853\\
\hline
OH&1085&1343\\
\hline
$\mathrm{H_2}$&387&479\\
\hline
$\mathrm{O_2}$&1042&1290\\
\hline
$\mathrm{H_2O}$&1601&1982\\
\hline
CO&1042&1290\\
\hline
HCO&1300&1609\\
\hline
$\mathrm{H_2CO}$&1515&1876\\
\hline
$\mathrm{CH_3O}$&1868&2313\\
\hline
$\mathrm{CH_3OH}$&1774&2195\\
\hline
$\mathrm{CO_2}$&2153&2664\\
\hline
\end{tabular}
\label{table-2}
\end{table}

\begin{table}[h]
\centering
\caption{Energy barriers against diffusion and desorption in K for
the olivine grain (See text for references.)}
\begin{tabular}{|l|l|l|}
\hline
Species &$E_b$ &$E_d$ \\
\hline
H&100&350\\
\hline
O&240&800\\
\hline
OH&378&1260\\
\hline
$\mathrm{H_2}$&135&450\\
\hline
$\mathrm{O_2}$&363&1210\\
\hline
$\mathrm{H_2O}$&558&1860\\
\hline
CO&363&1210\\
\hline
HCO&453&1510\\
\hline
$\mathrm{H_2CO}$&528&1760\\
\hline
$\mathrm{CH_3O}$&651&2170\\
\hline
$\mathrm{CH_3OH}$&618&2060\\
\hline
CO$_2$&750&2500\\
\hline
\end{tabular}
\label{table-3}
\end{table}


\section{Simulation procedure}
\begin{figure}
\vskip 0.6cm
\centering{
\includegraphics[width=5cm]{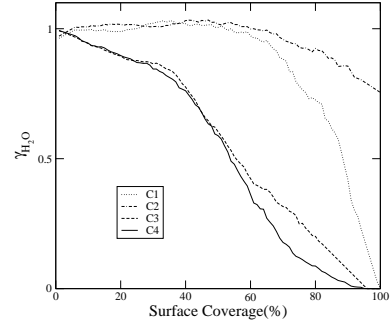}}
\caption{Variations in $\gamma_{H_2O}(t)$ with surface coverage 
for different reaction schemes for Model 2. Note that when the Eley-Rideal scheme
is included (C1 and C4) the production efficiency increases rapidly as the
grain is filled up.}
\end{figure}

We used a Monte-Carlo simulation method of studing the chemical evolution on grains
in the presence of H, O, and CO as accreting species. 
For the sake of simplicity we assume that a grain is square shaped having $S$ 
number of sites. For concreteness we chose olivine grains having
the surface density of sites $s=1.5 \times 10^{15}$ cm$^{-2}$ (Hasegawa et al. 1992).
We assumed further that each site has four nearest neighbors, 
as in an fcc[100] plane. To mimic the spherical 
grain structure, we assumed a periodic boundary condition, i.e., a species leaving the boundary site
of a grain on one side enters back into the grain from the opposite side.

In our method of simulation,
self-crossing paths of randomly walking atom are automatically included (some times 
called the back-diffusion in the literature, e.g., Chang et al. 2005). The minimum time 
step of our simulation is assumed to be the time taken for hopping or tunneling 
(whichever is smaller) of a H atom from site to site. If $r_{acc}$ is 
the astrophysically relevant accretion rate under consideration, 
the $i$-th species will have to be dropped after every $1/r_{acc}$ seconds. 
The location of the accreting $i$-th species is obtained from a 
pair of random numbers ($R_x, R_y$), themselves obtained
by a random number generator. This pair would place the incoming species 
at ($j$, $k$)-th site of the grain, where $j$ and $k$ are the nearest integers 
obtained using Int function: $j=int(R_x*n+0.5)$ and $k=int(R_y*n+0.5)$. Here, $n = \sqrt S$, 
being the number of sites on a square. Now during the hopping or tunneling
process, when one species enters the site that is already occupied by another reactant surface 
species, it will form a molecule if activation barrier energy permits.
Thermal evaporation of a species is also handled by random numbers.
After each hopping and tunneling time, we generate a random number ($R_t$) 
for $i$-th species, if $R_t$ is less than $W_i/a_{h_i}$ or $W_i/a_{t_i}$. Then we allow 
that species to evaporate.  Here, $W_i$ is the desorption rate and
$a_{h_i}$ and $a_{t_i}$ are the rates of hopping and tunneling 
of the  $i$-th species, respectively. We carry out our simulation up to the 
time within which one monolayer is produced. We wish to make a remark in passing  
that a totally different random number generator was also found to yield a similar 
result, so thus we believe that the results we present in this paper do not
depend on a random number generator.
\begin{figure}
\vskip 0.6cm
\centering{
\includegraphics[width=6cm]{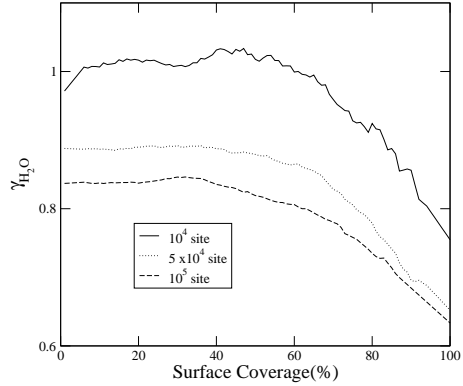}}
\caption{Variation in $\gamma_{H_2O}(t)$ with the surface coverage for different grain sizes.
Here, the binding energies of Model 2 are used. Note that $\gamma_{H_2O}(t)$ is decreased
with increasing grain size since the probability of capturing an incoming species 
increases with the grain size. Here C2 scheme has been chosen.}
\end{figure}

Keeping in mind that there are two types of reaction schemes, the landing of a gas phase 
atom or molecule can have one of these two distinct fates. It can land either 
on a vacant site or on an occupied site. When the landing is on a vacant site, 
the follow-up computation is straightforward. The species will simply start scanning the surface for a 
reactive partner. When it lands on an occupied site, however, there could be
two possibilities, (i) The incoming species can land directly on a reactive species
but it may or may not combine with it. If not, it will look for an alternate 
site for itself. (ii) The incoming species can land directly on a nonreactive 
species and look for an alternate site. However, looking for an alternate 
site itself could be achieved in two ways: in the same time step (which is more likely) or in the next 
time step. Thus we can have a total of four different models in the simulation. 
These are illustrated in a cartoon diagram C1-C4 in Fig.~\ref{fig-1}. In C1, both the LH 
and ER schemes are considered, i.e., the incoming atom/molecule is allowed to land on both 
the unoccupied and occupied sites. With the reactive species, it combines to form a 
new species (marked with a common circle). It skips the nonreactive species and looks for a
new site in the same time step. C4 is otherwise the same as C1, but the incoming 
species looks for a new site at the next time step after it lands on a 
nonreactive species. In C2, the incoming species lands on a reactive 
species, but does not combine and insteads finds a new site at the same time 
step (LH mechanism). C3 is the same as C2, but we allow the landing at a new time step. 

\begin{table}[h]
\centering
\caption{Surface Reactions in the H, O, and CO model}
\begin{tabular}{|l|l|l|}
\hline
Number & Reactions & E$_a$(K)              \\ 
\hline
1  & H+H $\rightarrow$ H$_2$          &      \\  
2  & H+O $\rightarrow$ OH             &      \\ 
3  & H+OH $\rightarrow$ H$_2$O        &      \\ 
4  & H+CO $\rightarrow$ HCO           & 2000 \\  
5  & H+HCO $\rightarrow$ H$_2$CO      &      \\ 
6  & H+H$_2$CO $\rightarrow$ H$_3$CO  & 2000 \\ 
7  & H+H$_3$CO $\rightarrow$ CH$_3$OH &      \\ 
8  & O+O $\rightarrow$ O$_2$          &      \\ 
9 & O+CO $\rightarrow$ CO$_2$        & 1000 \\ 
10 & O+HCO $\rightarrow$ CO$_2$+H     &      \\ 
\hline
\end{tabular}
\label{table-4}
\end{table}


\section {Model simulations}

We consider that only three species, namely, H, O, and CO are accreting in the grain phase,
and we assume that ten types of reactions can go on among these constituents. 
These are listed in Table~\ref{table-4}. Let us consider a reaction of the type,
\begin{equation}
a + b \longrightarrow c.
\end{equation}
The rate of variation of any molecular species $c$ with time is usually taken as
(see, Paper-I and references therein)
\begin{equation}
\frac{dn_c}{dt}=r_{ab}n_an_b/S 
\label{EQ-mn}
\end{equation}
where, $n_a$, $n_b$, and $n_c$ are the concentrations of the species $a$, $b$,and $c$
on a grain. Here, $r_{ab}$ is the rate of formation of $c$ out of $a$ and $b$,
and $S$ is the number of sites on the grain. The rate $r_{ab}/S$ is approximately the 
inverse of the time required for an atom to visit nearly all the adsorption sites on the grain surface. This
is because, in two dimensions, the number of distinct sites visited by a random walker is linearly 
proportional to the number of steps, up to a logarithmic correction. But in a realistic situation, 
this is not the case for two reasons (Paper-I). The first  
is that the number of hoppings to find a reactant partner depends on the number of atoms or molecules already occupying
the grain. When the grain is almost empty, a species has to hop several times, 
and when the grain is full, it requires less of hops, so the `size'
of a grain is relative as far as the hoping element is concerned. The second reason is the 
blocking effect. This phenomenon is seen very often on a grain having a mono-layer. 
On a grain surface, the interaction of species is generally considered to be
with those located in the four nearest neighboring sites. A possible scenario 
would be that a reactive species is blocked in all the four directions 
due to the presence of various species with which no reaction is allowed. 
For instance, if a CO molecule lands on a site surrounded by H$_2$O molecules,
it cannot react and will wait there itself for its desorption unless
the blocked species themselves desorb and free the sites for the landed species to move around. 
The effect becomes more severe as the grains start getting filled up. In the latter case,
the blocking effect will weaken when the blocked element overcomes the energy barrier. 
As the surface is populated, these types of effects become more and more prominent. 
To understand the influence of these processes on the recombination, we define two 
parameters, namely, $\gamma$ and $\alpha$ as given below.
\begin {figure}
\vskip 0.6cm
\centering{
\includegraphics[width=6cm]{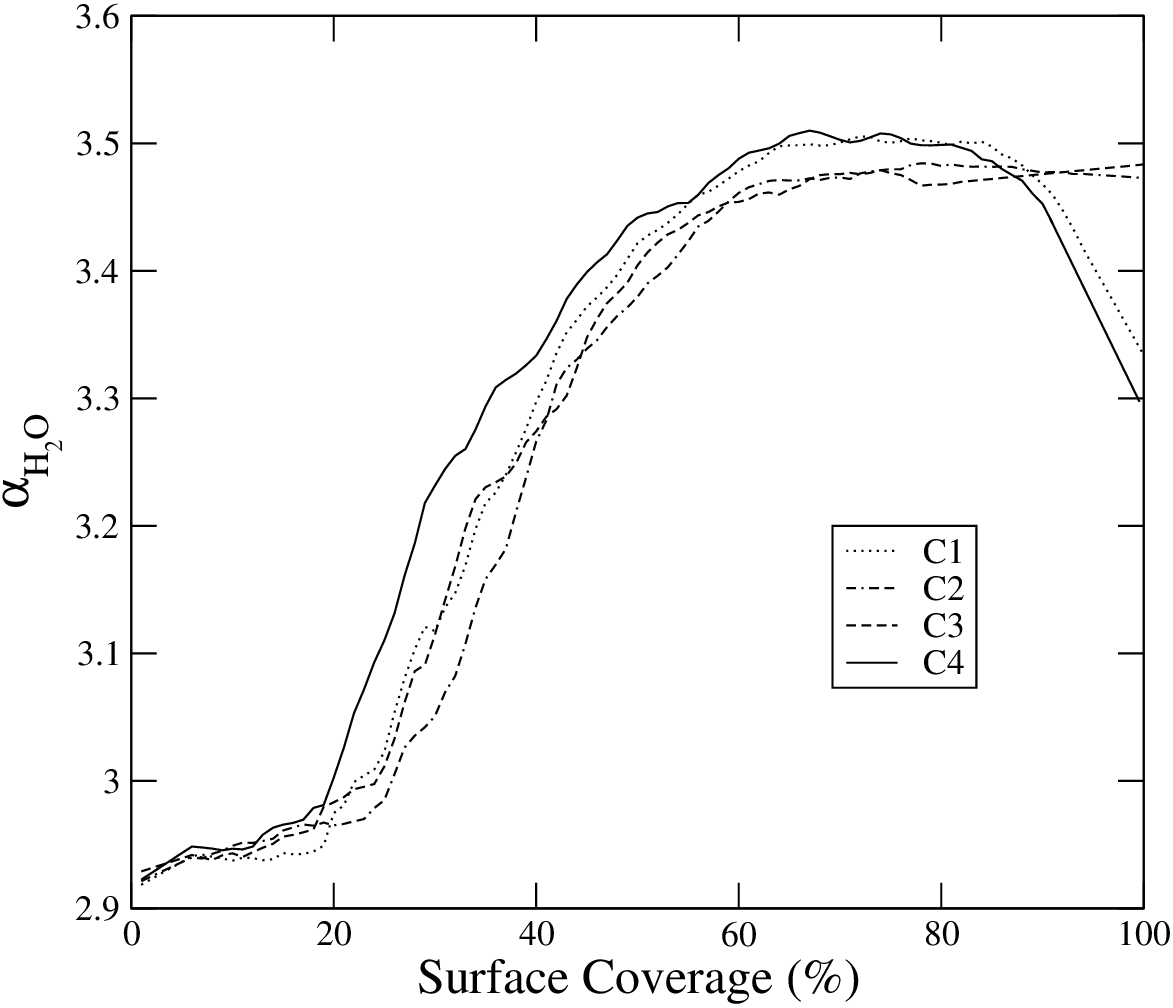}}
\caption{Variation in $\alpha_{H_2O}(t)$ with surface coverage for 
different reaction schemes as in Fig. 7. 
The formation rate of the new 
species goes down (i.e., $\alpha_{H_2O}(t)$ goes up) 
for C2 and C3 since the species is created in the next time step. }
\end {figure}

If $t_c$ is the ``average time" required to form a species $c$ after the species $a$ and $b$ 
are simultaneously present in our simulation, then $\frac{dn_c}{dt} \sim n_a n_b/t_c$, where $t_c 
= r_{ab}/S$.  The rate with which the species $c$ is formed is $1/t_c$. Because 
of the two effects mentioned above, the actual rate may be assumed to be 
$r_{ab}/S^\gamma_c$ instead of $r_{ab}/S$. From the above equation, 
\begin{equation}
\gamma_c=\frac{log[r_{ab}t_c]}{log S}.
\end{equation}
As an example, let us consider the calculation of $\gamma_c$ for H$_2$O. 
Since H has a very small desorption time scale compared to the other species, 
it will desorb from the grain if it does not find a reaction partner within 
its desorption time scale. If H gets another OH to form H$_2$O, then we 
calculate $dt_{H_2O}$ by noting down the time required to form an 
H$_2$O after the accretion of an H. 
At the beginning of the simulation, most of the sites are empty and one H will have 
to more or less scan the whole of the grain to find its reactant partner. 
and hence will take longer time to form a molecule. On the contrary, 
when most of the sites are filled with the different species, it will take less time 
to form a molecule and $\gamma_c$ should decrease. 
When a molecule is formed due to direct hitting, i.e., through the ER mechanism, 
$\gamma_c$ is clearly zero, since production rate is independent of the 
population of the species on the grain. Thus, if ER mechanism is included,
$\gamma_c$ tends to become negligible as the grain is populated.

\begin {figure}
\vskip 0.6cm
\centering{
\includegraphics[width=6cm]{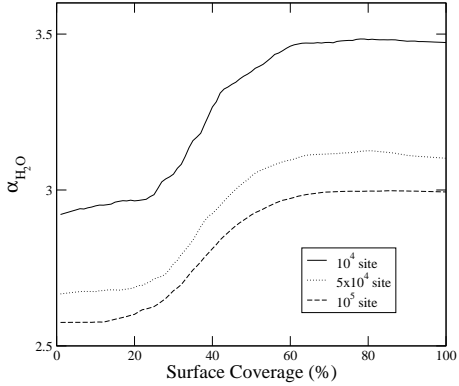}}
\caption{Same as in Fig. 8 except that $\alpha_{H_2O}(t)$ has been plotted. As the grain
size gets smaller, $\alpha_{H_2O}$ increases, i.e., the recombination efficiency decreases. 
See text for details. }
\end {figure}

While $\gamma$ is derived from the time interval between the release of the reactants
and the actual reaction to take place, another quantity, namely $\alpha$ may be defined
to consider the cumulative effects and thus depends on the total time taken
since the start of the accretion onto the grain.
Because of the two competing processes mentioned at the beginning of this section,
there will be a net change in the rate equation itself with time, since $\alpha$ itself
will be time dependent. To quantify this, we replace $r_{ab}/S$ in Eq.~\ref{EQ-mn}
by $r_{ab}/S^{\alpha(t)}$. Thus, the Eq.~\ref{EQ-mn}, which is rewritten as
\begin{equation}
log{\frac{dn_c}{dt}}=log[r_{ab}n_an_b/S^{\alpha_c(t)}],
\end{equation}
can be used to obtain $\alpha_c(t)$ as
\begin{equation}
\alpha_c(t)=log[\frac {(r_{ab} n_a n_b)}{\frac {d n_c} {d t}}]/log(S).
\label{EQ-al}
\end{equation}
Note that, for reactions between the similar species, their will have to be a factor of 
$1/2$ inside the logarithmic expression (see, Paper-I for H$_2$ formation)
to reduce the possibility of double counting.
As the surface is populated by a number of reactant species, the recombination time scale
would vary with time and thus $\alpha_c(t)$ will also vary. As mentioned before, the
time $t$ used in computing $\alpha$ is the total time taken since the grain starts populating. 
On the other hand, $\gamma$ is an average quantity that is computed instantaneously. 

We run the simulations to show that $\alpha(t)$ and $\gamma$ are not necessarily unity. 
This means that the surface chemistry is not just a matter of random hopping, since 
the degree of randomness is seriously modified with the population and 
the abundance of species on the grain surface. The Monte-Carlo simulation
reflects a more realistic computation, but it is very time-consuming.
In a large-scale computation, it is impractical to do such a simulation. Indeed,
we also show that only when $\alpha$ as given above is used in the 
effective rate equation (where every $S$ is replaced by $S^\alpha_i$, where $\alpha_i$
is different for different species), then
the abundances of each species obtained from such an equation agree with the results of
the Monte-Carlo simulation. Thus, one may as well solve the effective
rate equation using $\alpha$ obtained from our calculations. We have made
a table for $\alpha$ (Table 7 , below) for future. However,
discussions of its approximate value, shape, and dependence is beyond the
scope of this paper and will be discussed elsewhere.

In what follows, we ran twelve different cases. 
For each set of binding energies, all the four C1, C2, C3, and C4 
cases are considered. Due to the complexity of the system, we restricted 
ourselves only up to the formation of a mono-layer on the grain surface.

\section{Results} 
\begin {figure}
\vskip 0.6cm
\centering{
\includegraphics[width=7cm]{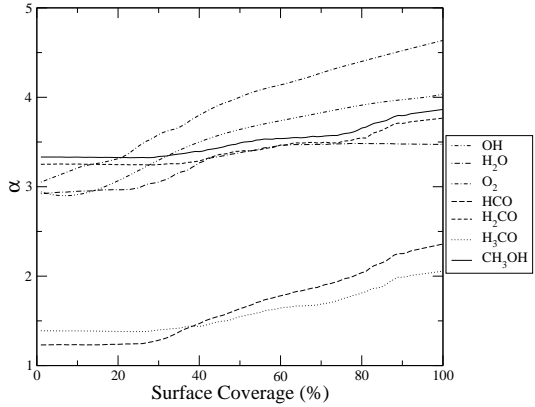}}
\caption{Variation in $\alpha$ with the surface coverage for various molecular
species that form on the grain surface. All of them deviate from unity.
This case is for Model 2 with the C2 scheme.} 
\end {figure}

We first carried out the simulation for a grain having $10^4$ sites. We  considered two different 
sets of gas-phase abundances in our study, one with a high density and the other with low density. 
For each set of gas-phase abundances, we consider three sets of binding energies and 
for each set of binding energies we have four cases, C1, C2, C3, and C4 as 
described in Section 3. Thus for each set of gas-phase abundance, we essentially have 
twelve different cases. 

\subsection{Grain surface abundances}
\begin {figure}
\vskip 0.6cm
\centering{
\includegraphics[width=6cm]{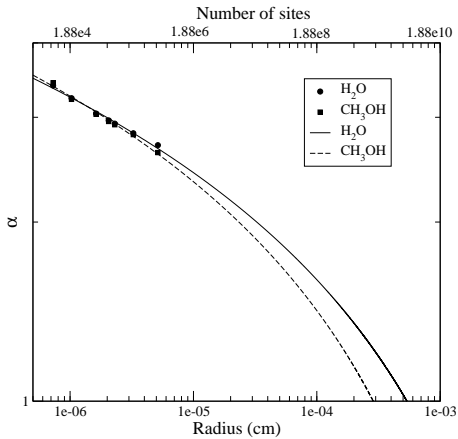}}
\caption{Variations in $\alpha$ for water and methanol (averaged over the surface coverage)
as functions of the grain radius and no. of sites on the grain. The filled circles
are for $H_2O$ and the filled boxes are for methanol. The corresponding fitted curves are 
also shown. They are extrapolated to larger grain sizes.} 
\end {figure}

In Fig.~\ref{fig-2}, we show the results for the high gas-phase abundances. In 
the first column, the plots 
correspond to the binding energies of Model 1 as mentioned in the 
Table~\ref{table-2} are shown.
The binding energies are high and consequently the diffusion rates are low. 
Let us first consider C1 and C2 cases. We note that, respectively, 
only $16 \%$ and $11 \%$ of the grain deposition is water. This low production 
of water is due to the slow diffusion rate and due to the absence of sufficient 
number of hydrogen atoms to react with the oxygen atom. In C1, the excess of production of 
water by four per cent is mainly due to ER mechanism. There is hardly any methanol 
produced due to slow diffusion rate and due to presence of an activation barrier energy. 
In C1, around $19 \%$ of molecular oxygen is produced, but its absence in C2 suggests 
that it is produced due to the ER mechanism. Around $50 \%$ of the grain surface abundance is CO. 
The time required to grow one mono layer for C1 and C2 is $1600$ years and $1300$ years,
respectively. The grain is very quickly filled up by the accreted species. The results for the 
cases C3 and C4 differ only slightly in comparison to C2 and C1, respectively.
Only a small increase in the molecular species and decrease in the atomic species is observed.
The times required to grow a mono layer in C3 and C4 are around $13000$ 
years and $36000$ years, respectively. This
longer timescale is due to the fact that in these cases atoms/molecules are not 
bound to accrete on the same time step. In the case of C4 we continue our 
simulation up to the time when all O and OH are converted to H$_2$O, hence the time taken is much longer. 

The second and third columns show the plots corresponding to binding energies 
as mentioned in Table~\ref{table-3}. In the second column, the hydrogen 
diffusion is due to the thermal hopping, whereas in the third column 
this is due to tunneling. These binding energies are low and the diffusion rates are consequently high. 
In all the four cases, the most dominant species are CO, O$_2$, and water.
Water abundance is very similar, varying between $11 \%$ and $19\%$. A very small amount of
methanol has been produced. In all the cases, the production of a significant amount 
of molecular oxygen suggests that the ER mechanism may have an active role to play. The time required to grow a 
mono-layer is around $1500$ years.  The third column also shows a similar trend of the results.
\begin {figure}
\vskip 0.6cm
\centering{
\includegraphics[width=6cm]{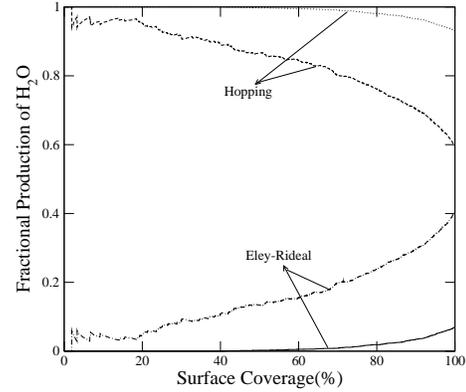}}
\caption{Variation in the fractional productions of H$_2$O through ER and LH schemes
against the surface coverage of a mono-layer. For low surface coverage, H$_2$O 
production is dominated by the hopping process and when surface coverage is higher, 
the Eley-Rideal mechanism contributes in a major way towards the H$_2$O 
production. The uppermost and the lowermost curves are for the Model 2 energy and
C1 scheme, with  the low abundances of the accreting species.
We compare the same for Model 1 energy with C1 method, 
with the high abundances for the accreting species (middle two curves).}
\label{fig-10}
\end {figure}

In Fig.~\ref{fig-3}, we show the results for low gas-phase abundance. A significant amount of water 
is produced in all the cases, since the H/O and H/CO is relatively high.
The presence of the atomic oxygen and carbon monoxide is also much lower, especially for
Models  2 and 3. This is because, in the last cases the mobility of H being
many orders of magnitude higher hydrogenation process is more efficient and that uses up
O and CO to create water and methanol.
This makes the production of methanol significant for Models 2 and 3 and 
insignificant for Model 1.
Compared to the high abundance cases (Fig. 2), the accretion rate for 
low abundance is much lower, so 
the grain is not has full of atomic species as before. The time required to grow 
a mono-layer is around $0.1$Myr for both the cases. A similar trend is found for C3 and C4, 
but it takes around $1$Myr to fill up the grain. The second and third columns of Fig. 3 
show a similar result. In all the cases, the most dominant species is water. 
The abundance of water varies between $46\%$ and $61 \%$. The next dominant species 
is methanol with an abundance between $26 \%$ and $33 \%$. But only a trace amount of molecular 
oxygen is produced. Table~\ref{table-5} shows the percentage abundances for selected
species, and Table~\ref{table-6} shows the time required to grow a mono-layer.

\begin {figure}
\vskip 0.6cm
\centering{
\includegraphics[width=6cm]{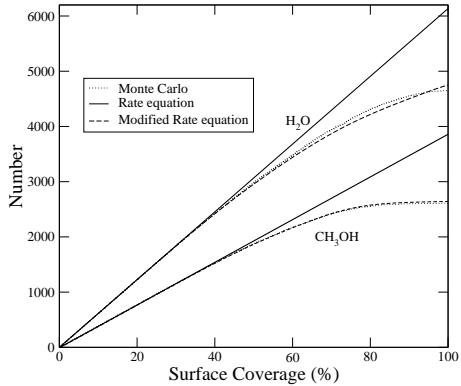}}
\caption{Comparisons of the number of H$_2$O and CH$_3$OH on a grain of $10^4$ sites
as obtained by the rate equation method, the Monte-Carlo method, and 
the effective rate equation method. Here we chose the Model 2 energies 
and  C2 with the low abundances case. The agreement between the 
Monte-Carlo result and the effective rate equation result is very good.}
\label{fig-14}
\end {figure}

Figure 4(a-b) shows the rate  $R_{H_2O}$ of production of H$_2$O for various mechanisms. This has been 
computed in two ways. Figure 4a shows the cumulative average rate where $R_{H_2O} = $H$_2$O$(t) /t$. 
We clearly see that, unless the production rate is very high (e.g. in case of C1), 
the cumulative rate generally falls off very rapidly with time. 
In the C1 scheme, the abundance of H$_2$O is higher because the formation 
of H$_2$O is dominated by direct hitting rather than hopping. In Fig. 4b, 
we plot $R_{H_2O}=\frac{dH_2O}{dt}$, where $d$H$_2$O is the production of H$_2$O 
in $dt$ time in which a fixed percentage $dp\%$ of the surface coverage is increased.
Thus, in effect, it is the instantaneous rate. For simplicity, we choose $dp=10\%$ 
so as to get smooth curves. In the case of the LH scheme, i.e., for C2 and C3, 
as the surface starts to populate, the rejection of the incoming species become important and thus the
rate of production decreases with the surface coverage. 

\subsection{Temperature dependence}

Figure 5 shows the production of various species 
as a function of surface coverage at $10$ K, $15$ K and
$20$ K. We carried out this computation only for Model 2 
and taking low abundances of the accreting species.
In all the three temperatures, the water is the most abundant species. But CH$_3$OH is only efficiently
formed in $10$ K and $15$ K. At $20$ K, methanol production is significantly low because the desorption 
time scale of atomic hydrogen is much shorter and, at the same time, the activation barrier
energy is much higher to produce methanol. It is instructive to study the time taken to fill up 
the grain. In Fig. 6 we plot the time required to build a mono-layer on a grain of $10^4$ sites as a 
function of the temperature of the grain. At higher temperatures, the time taken is 
lower because the grain is full of nonreactive CO (Fig. 6). 

\subsection{Variation in $\gamma$ and $\alpha$}
In Fig. 7, the variation in `catalytic capacity' for water, namely, $\gamma_{H_2O}(t)$ (Eq. 4) is shown
as a function of the surface coverage. The figure shows that 
$\gamma_{H_2O}(t)$ gradually decreases with the surface coverage. This is because
an increase in the surface coverage will decrease the number of vacant sites and will decrease the time 
for searching its reactant partner to recombine. The blocking 
effect also shows its effect by reducing $\gamma_{H_2O}$ very 
rapidly. The lower the value of $\gamma_{H_2O}$, the more efficient the recombination. This parameter is 
dependent on the number of sites on the grain. Figure 8 shows the change of $\gamma_{H_2O}$ with 
the number of grain sites. Naively, one would have expected that $\gamma_{H_2O}$ to remain 
unity always. In that case, $\gamma_{H_2O}$ would be independent of number of sites in the grain.
However, we find that $\gamma_{H_2O}$ goes down with increase in $S$. From Eq. (4) we note that, 
when $S$ is increased from $10^4$ to $10^5$, the value of $\gamma_{H_2O}$ should go down 
by about $20\%$. This is roughly what is seen in the figure.

In Fig. 9, the variation in $\alpha_{H_2O}$ (Eq. 6) for water is 
shown as a function of time, which is represented 
here as the surface coverage. Here $10^4$ sites are chosen. This exponent deviates significantly from $1$. This 
deviation is mainly due to blocking the reacting species by the nonreactive species and the 
population of various species on the grain surface. This is particularly true for cases 
C2 and C3, since the ER mechanism was not assumed. For those with an ER mechanism (C1 and C4), 
the numbers actually go down, so the formation rate goes up. In Fig. 10, 
we plot the variation in $\alpha_{H_20}$ as a function of surface coverage for a different number of sites 
on the grain. As the number of sites increases, $\alpha_{H_2O}$ is seen to 
decrease as in the case of $\gamma_{H_2O}$. We thus observe 
that the production rate of any species goes up for larger
grains. To judge the implication of this in a cloud, we note that the dependence 
of the number density on the radius $r$ goes typically as $r^{-3.5}$ (Mathis et al. 1977), while
the dependence of the production rate is $r^n$ with $n \sim 1$. Thus the production 
is still dominated by the smaller grains. However, smaller grains are lesser stable and could easily
evaporate. This aspect clearly requires more thorough study. 

In Fig. 11, we plot $\alpha$ for various species as a function of surface coverage when $10^4$ sites and the 
Model 2 with C2 mechanism were chosen. The most important observation from this is that $\alpha$ is not 
unity for any of these species. Second, up to about $30\%$ of the surface coverage $\alpha$ roughly remains
constant, but after that all of them are going up monotonically as the surface gets filled up.
Here the ER scheme is not used, and the formation rate goes down with coverage.

Since we generally see that $\alpha$ deviates from unity for smaller grains, it is 
pertinent to ask at, roughly what grain sizes $\alpha$ becomes unity. This would 
give us a region in which the classical rate equation would be valid. In Fig. 12,
we show the variations in $\alpha$ for water and methanol 
are shown as functions of the grain radius and number of sites on the grain. 
The corresponding fitted curves are
also shown. They are extrapolated to larger grain sizes. 
We note that, $\alpha \sim 1$ at around $r \sim 3-5\mu$m.
Since the computational time to actually carry out the simulation at this
size is prohibitively high, it could not be independently verified if this conclusion is 
rigorously valid. On the other hand, the number of grains at this size is expected to be
negligible, and thus the exact knowledge of this limit may not be essential.  

\subsection{A comparison between LH and ER schemes}
\begin {figure}
\vskip 0.6cm
\centering{
\includegraphics[width=6cm]{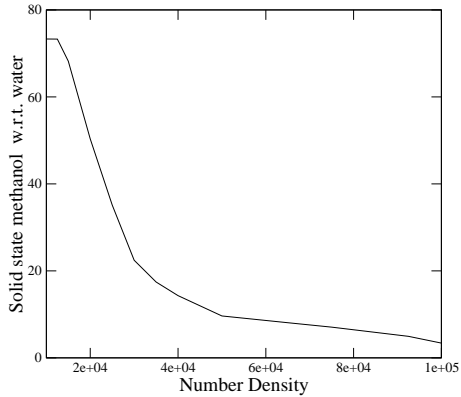}}
\caption{The ratio of methanol and water abundances on the grain 
surface plotted against the number density of the accreting gas. 
From the figure it is clear that as the number density goes up, methanol
production goes down. The observed abundance of solid state methanol w.r.t. water 
is 5\%-30\%. From the figure it is to be noted that
the region lies in the range $2.6 \times 10^4$ to $8 \times 10^4$.
These simulation were carried out for Model 2 by considering the C2 method.}
\label{fig-15}
\end {figure}

In Fig. 13, we plot the fractional production of H$_2$O for two different
schemes at a given surface coverage to show the relative importance between
the ER and LH schemes. Initially, when most of the sites are empty, the production 
of H$_2$O is mainly due to the hopping mechanism i.e., the LH scheme. As the 
grain starts to populate, the ER scheme starts to play an important role. 
In the figure, the uppermost and the lowermost curves are obtained using 
the low abundance set with C1 model. The other pair of curves is for 
a higher abundance set in the gas phase. Here the effect 
is more prominent. In general, the abundance of all the species 
is always a few $\%$ higher if the ER scheme is allowed. 

\subsection{Comparison of results with the effective rate equations}

It may be instructive to check if the rate equations are modified
with $r_{ab}/S$ replaced by $r_{ab}/S^\alpha$ everywhere, and they 
are solved by the usual method (Acharyya et al. 2005), then whether the 
results become comparable to those obtained from our Monte-Carlo
method. In Fig. 14, we made a comparison of the abundances of two of the important 
species, namely, H$_2$O and CH$_3$OH in different methods. The results of our simulation
clearly agree with those from the {\it effective} rate equation very well, but not 
with the original rate equation. This indicates that the $\alpha$s we introduced are 
very important and their effects are to be taken into account for accurately
estimating of the species from the grain chemistry. Indeed, using the well-known
MRN model of grain size, we find that water 
and methanol will be underproduced by $\sim 25\%$ and $\sim 33\%$, respectively,
when the effective rate equations are used instead of the usual rate equation.  

Since $\alpha$ depends on both the gas and the grain parameters in a non-trivial way,
it is difficult to obtain any analytical formula or fitted curves without repeating
the simulation many times. This is beyond the scope of this paper and will be reported
in future. Nevertheless, we provide in table 7, a list of $\alpha$ for olivine grains 
at $10$K for Model 2 binding energies and C2 scheme of interaction as a function of the
surface coverage and number of sites. For other sites and coverages, one can use 
an interpolation method. We do not give any table for $\gamma$ since it is not immediately
used in the rate equation.

\section{Concluding remarks}

In this paper, we studied the formation of water, methanol, and other related species on a grain surface 
using a Monte-Carlo method. We used three different sets of binding energies, two different gas-phase 
abundances and considered both the ER and LH mechanisms. We considered all the four possible ways (C1-C4) 
of the landing of a species on the occupied site of a grain. Besides the formation of these species, we 
also calculated two parameters $\gamma$ and $\alpha$ that represent the recombination efficiencies 
to show that they are indeed dependent on the number of sites on the grain and the populations of various
species on a grain surface.

We found that the formation of various molecules is dependent on the binding energies. We find that, 
when the higher 
binding energies are used, it is very difficult to produce a significant amount of the molecular  
species, instead, the grain is found to be full of atomic species. The formation 
of these species is also dependent on the gas phase density. We found that, for the high density case, 
the production of various molecules is small and the grains are 
filled up relatively quickly by atomic species.
We found that, if both the ER and LH mechanisms are considered, then the production is always high
and the grain is filled up very quickly. As expected, we find that, when the grain is more or less empty,
the LH scheme is most important. The ER scheme starts dominating more and more as the grains are filled up.
We have verified that our results from the Monte-Carlo simulation match with that from the effective rate equation. 

It is important to see if we can put a constraint on the model parameters from the observational 
results. We already mentioned in the Introduction that the abundance of 
water relative to H$_2$ in cold clouds is between $10^{-10}$ and $8 \times 10^{-10}$ and
between $10^{-6}$ to $10^{-4}$ in hot cores. The abundance of methanol 
with respect to H$_2$ is $ \sim 10^{-9}$ for coldest clouds and between $\sim 10^{-9}$ 
to $10^{-7}$ for warmer clouds and $\sim$ a few $\times 10^{-8}$ in hot cores. In the grain 
surfaces, the solid state methanol abundance should be at the most $30\%$ with respect to H$_2$O. 
If we glance at Table 5 and assume that the surface coverage on multiple layered grains is similar 
to that on a mono-layer, then it is clear that the case of Model 1 and high abundance cases for Models 2 and 3
are not very relevant for the production of methanol. Indeed, the low abundance cases with 
Models 2 and 3 produce a similar solid state methanol with respect to H$_2$O ($\sim 43\%$ - $\sim 57\%$). 
Corresponding high abundance models produce only $4\% - 6\%$. Thus the observed 
abundance of $\sim 30\%$ must be from some intermediate region of the cloud. 
In Fig. 15 we show the ratio of methanol and water as a function of the
number density of the gas phase, i.e., accretion rate. 
We clearly see that to obtain the observed ratio the number density should be
around $2.6-8 \times 10^4$cm$^{-3}$.

If the temperature is high enough, some methanol is released 
in the gas phase. If $n_l$ layers were formed on the grains 
and were all assumed to evaporate to the gas phase, the gas phase abundances of 
methanol would be given by
$$
R = \frac{n_g n_l s S}{n_{H_2}} =3 \times 10^{-6} f_{-12} n_{10} s_{i,0.3} S_6 .
$$
Here, $n_g \sim n_{H_2} 1.33 \times 10^{-12}=n_{H_2} f_{-12}$ the number density of grains, 
$n_{H_{2}}$ is the number density of H$_2$ in the gas phase, $s_{i,0.3}$ the fractional 
surface coverage of $i$th species (e.g., methanol or water) in units of  $0.3$, $S_6$ the number of sites in the 
grain in units of $10^6$, and $n_{10}$ in units of $10$ mono-layers. 
Since time taken for production of a mono-layer is $ \sim 0.1$ Myr, by the time $10$ 
mono-layers are produced, the grains are deeply inside in higher abundance region. 
Here the mono-layer production time is lower but the production efficiency of 
formation of methanol is lower as well. Thus $R$ computed for methanol 
above, for $n_l=1$ to $10$ varies from $\sim 10^{-7}$ ($s_{i,0.3} \sim 0.03$) to $3 \times 10^{-5}$
($s_{i,0.3} \sim 1$), generally
agrees with the observational results. If, for instance, a fraction of grains are sublimated,
say, $f_{-12} \sim 0.1$, the abundance could be even lower. 
For water, $R\sim 10^{-6}$ to $6 \times 10^{-5}$, where we have put
$n_{10}=1$ to $10$, $s_{i,0.3}=0.33$ for high abundance, and $s_{i,0.3}=2$ 
for low abundance. This is also in the observed range.

One of our important findings is that the parameter $\alpha(t)$ strongly depends on the 
population of the reactant species on the grain surface. This deviates significantly from unity.
This seems to be a very important parameter, because in the usual rate equation we assume 
that $\alpha(t)$ is always $1$ (a consequence of the assumption that the recombination 
is totally a random walk process). This is an overestimation. In Paper-I we also computed this
parameter for the H$_2$ molecule. We defined another parameter $\gamma$ called 
the `catalytic capacity' and found that it goes down with the increased surface population. 
This shows that the rate of production indeed increases as the grain is filling up. We also found 
that the behaviors of $\alpha(t)$ and $\gamma$ strongly depends upon the grain temperature.

In our present calculation we restricted ourselves in two ways: (a) We considered only dense
clouds where the accretion is such that most of the hydrogen is already in the molecular form
in the gas phase. Thus accreting gas composition produces primarily water and methanol as is truly
the case for dense clouds. In diffused clouds, on the other hand, the  composition is such that
mostly molecular hydrogen is formed and are desorbed into the gas phase. Such a work was
presented in Paper-I. (b) We restricted ourselves up to the formation of a mono-layer 
due to the fact that the complexity of the problem rises with layer number and the unavailability 
of sufficient data (e.g., the binding energies at different surfaces). In reality, multi-layers
would be produced and each layer is expected to have a different abundance due to the freezing out effect.
For instance, in the first layers would be dominated by H$_2$O and methanol, etc. In the later stage,
accretion species will be dominated by CO, O$_2$, N$_2$, etc. Detailed results are in progress
and will be reported elsewhere.

The paper has been greatly improved due to the helpful comments of the anonymous referee  
who is acknowledged. The work of AD was supported by a RESPOND grant from ISRO. 

{}

\begin{table}
\scriptsize
\caption{Surface coverage of the major species when one mono layer is built}
\begin{tabular}{|c|c|c|c|c|c|c|c|c|c|c|c|c|c|c|}
\hline
Major&Used&\multicolumn{4}{|c|}{$E_b$ \& $E_d$ from Table 2}
&\multicolumn{4}{|c|}{$E_b$ \& $E_d$ from Table 3}
&\multicolumn{4}{|c|}{$E_b$ \& $E_d$ from Table 3}\\
species&abundances&\multicolumn{4}{|c|}{Tunneling not allowed (Model 1)}
&\multicolumn{4}{|c|}{Tunneling not allowed (Model 2)}
&\multicolumn{4}{|c|}{Tunneling allowed (Model 3)}\\
\hline
&&C1&C2&C3&C4&C1&C2&C3&C4&C1&C2&C3&C4\\
\hline
O&high&4.98&32.97&33.29&0&1.8&13.13&12.74&0&1.72&13.17&12.83&0\\
\cline{2-14}
(in \%)&low&0&12.56&12.61&0&0&13.82&10.24&0&0&13.67&10.68&0\\
\hline
O$_2$&high&19.32&0&0&22.49&25.26&19.77&19.76&28.0&25.39&19.32&19.43&28.16\\
\cline{2-14}
(in \%)&low&0.24&0&0&0.38&0.67&0.39&0.76&1.05&0.56&0.35&0.56&0.6\\
\hline
OH&high&8.92&11.13&10.97&0&2.89&2.35&2.32&0&2.92&2.19&2.34&0\\
\cline{2-14}
(in \%)&low&0&0.2&0.22&0&0&0.15&0.27&0&0&0.15&0.2&0\\
\cline{2-14}
\hline
H$_2$O&high&16.33&11.15&11.19&32.97&13.36&10.57&11.23&18.82&13.23&11.34&11.17&18.84\\
\cline{2-14}
(in \%)&low&61.04&48.59&48.71&61.43&60.41&46.81&49.69&60.27&60.55&47.02&49.25&60.47\\
\hline
CO&high&50.45&44.75&44.54&44.54&48.60&44.99&45.38&44.19&48.61&45.42&45.69&42.47\\
\cline{2-14}
(in \%)&low&38.58&38.57&38.38&38.1&12.66&12.41&9.67&12.31&12.16&12.68&9.72&12.12\\
\hline
H$_2$CO&high&0&0&0&0&3.86&4.07&3.86&4.46&3.96&3.75&4.07&4.65\\
\cline{2-14}
(in \%)&low&0.13&0.07&0.07&0.09&0.16&0.15&0.15&0.15&0.17&0.12&0.12&0.16\\
\hline
CH$_3$OH&high&0&0&0&0&0.54&0.64&0.53&1.06&0.58&0.53&0.53&1.16\\
\cline{2-14}
(in \%)&low&0&0&0&0&26.01&26.16&29.03&26.14&26.41&25.87&27.2&26.26\\
\hline
CO$_2$&high&0&0&0&0&3.0&1.29&1.15&3.47&2.97&1.3&1.03&4.72\\
\cline{2-14}
(in \%)&low&0&0&0&0&0.08&0.06&0.11&0.08&0.14&0.05&2.18&0.39\\
\hline
\end{tabular}
\label{table-5}
\end{table}
\begin{table}
\scriptsize
\caption{Time taken to build one mono-layer}
\begin{tabular}{|c|c|c|c|c|c|c|c|c|c|c|c|c|c|c|}
\hline
&Used&\multicolumn{4}{|c|}{$E_b$ \& $E_d$ from Table 2}
&\multicolumn{4}{|c|}{$E_b$ \& $E_d$ from Table 3}
&\multicolumn{4}{|c|}{$E_b$ \& $E_d$ from Table 3}\\
&accretion &\multicolumn{4}{|c|}{Tunneling not allowed (Model 1)}
&\multicolumn{4}{|c|}{Tunneling not allowed (Model 2)}
&\multicolumn{4}{|c|}{Tunneling allowed (Model 3)}\\
\hline
&&C1&C2&C3&C4&C1&C2&C3&C4&C1&C2&C3&C4\\
\hline
Time&high&1.6(3)&1.3(3)&1.3(4)&3.6(4)&1.7(3)&1.6(3)&8.3(3)&3.2(4)&1.7(3)&1.6(3)&5.7(3)&3(4)\\
\cline{2-14}
in Year&low&1.1(5)&1.1(5)&9.4(5)&1.1(6)&1.1(5)&1.1(5)&2.4(5)&1.0(6)&1.1(5)&1.1(5)&3.2(5)&1(6)\\
\hline
\end{tabular}
\label{table-6}
\end{table}
\begin{table}
\scriptsize
\caption{Calculated value of $\alpha$, for different species, 
using low abundances of the accreting species for Model 2 and keeping 
the grain at 10K. These simulations were carried out for the C2 scheme
by considering an Olivine grain having $10^4$ number of sites.}
\begin{tabular}{|c|c|c|c|c|c|c|c|c|c|c|c|}
\hline
&&\multicolumn{10}{|c|}{Surface coverage in (\%)}\\
\hline
Species&Site&10&20&30&40&50&60&70&80&90&100\\
\hline\hline
&$10^4$&2.92&3.08&3.32&3.53&3.68&3.79&3.88&3.98&4.05&4.12\\
\cline{2-12}
OH&$5 \times 10^4$&2.71&2.78&2.98&3.12&3.23&3.32&3.4&3.47&3.53&3.59\\
\cline{2-12}
&$10^5$&2.61&2.69&2.83&2.98&3.09&3.18&3.25&3.31&3.37&3.42\\
\hline
&$10^4$&2.95&2.97&3.05&3.27&3.34&3.46&3.47&3.48&3.48&3.47\\
\cline{2-12}
H$_2$O&$5 \times 10^4$&2.67&2.69&2.76&2.92&3.04&3.1&3.12&3.13&3.11&3.1\\
\cline{2-12}
&$10^5$&2.58&2.6&2.68&2.81&2.92&2.97&2.994&2.997&2.997&2.994\\
\hline
&$10^4$&3.24&3.34&3.61&3.83&4.03&4.17&4.3&4.43&4.55&4.67\\
\cline{2-12}
O$_2$&$5 \times 10^4$&2.55&2.64&3.04&3.29&3.5&3.66&3.8&3.92&4.03&4.12\\
\cline{2-12}
&$10^5$&2.36&2.52&2.79&3.07&3.3&3.46&3.6&3.71&3.81&3.89\\
\hline
&$10^4$&1.23&1.24&1.28&1.47&1.64&1.78&1.89&2.04&2.25&2.36\\
\cline{2-12}
HCO&$5 \times 10^4$&1.08&1.14&1.29&1.43&1.55&1.68&1.78&1.89&2.04&2.18\\
\cline{2-12}
&$10^5$&1.07&1.11&1.25&1.41&1.54&1.64&1.74&1.85&1.98&2.14\\
\hline
&$10^4$&3.25&3.25&3.25&3.29&3.4&3.47&3.49&3.55&3.71&3.77\\
\cline{2-12}
H$_2$CO&$5 \times 10^4$&2.77&2.78&2.81&2.95&3.06&3.11&3.13&3.15&3.22&3.32\\
\cline{2-12}
&$10^5$&2.64&2.66&2.73&2.84&2.95&3.0&3.02&3.06&3.13&3.24\\
\hline
&$10^4$&1.39&1.38&1.4&1.44&1.55&1.64&1.69&1.81&1.99&2.06\\
\cline{2-12}
H$_3$CO&$5 \times 10^4$&1.2&1.21&1.24&1.37&1.49&1.57&1.63&1.69&1.79&1.89\\
\cline{2-12}
&$10^5$&1.17&1.19&1.26&1.37&1.47&1.54&1.59&1.65&1.75&1.88\\
\hline
&$10^4$&3.33&3.32&3.33&3.39&3.49&3.54&3.56&3.66&3.79&3.864\\
\cline{2-12}
CH$_3$OH&$5 \times 10^4$&2.86&2.87&2.9&3.01&3.12&3.18&3.21&3.24&3.33&3.43\\
\cline{2-12}
&$10^5$&2.71&2.72&2.78&2.87&3.0&3.04&3.08&3.11&3.19&3.31\\
\hline
\end{tabular}
\end{table}

\begin{thebibliography}{}
\def\ref#1\par{\parshape=2 0in 14.5cm 1cm 13.5cm {#1} \par}
\parskip=0pt
\parindent=0pt

\bibitem{} Acharyya, K., Chakrabarti, S.K. and Chakrabarti, S. 2005, {\it MNRAS}, {\bf 361}, 550
\bibitem{} Allen, M. and Robinson, G. W., 1975., {\it ApJ}, {\bf 195}, 81  
\bibitem{} Allen, M. and Robinson, G. W., 1976., {\it ApJ}, {\bf 207}, 745 
\bibitem{} Allen, M. and Robinson, G. W., 1977., {\it ApJ}, {\bf 212}, 396 
\bibitem{} Bachiller, R. and Perez Gutierrez, M.,1997, ApJ, 487L, 93
\bibitem{} Bachiller, R., Codella, C., Colomer, F., Liechti, S. and Walmsley, C. M., 1998, A\&A, 335, 266
\bibitem{} Biham, O., Furman, I., Pirronello, V. and Vidali, G., 2001, {\it ApJ} {\bf 553}, 595
\bibitem{} Boogert, A.C.A and Ehrenfreund, P.,2004, ASPC, 309, 547
\bibitem{} Chakrabarti, S.K., Das, A., Acharyya, K. and Chakrabarti, S., 
2006a, {\it A\&A}, {\bf 457}, 167 (Paper I)
\bibitem{} Chakrabarti, S.K., Das, A., Acharyya, K. and Chakrabarti, S., 2006b, {\it BASI}, {\bf 34}, 299
\bibitem{} Chang, Q., Cuppen, H. M. and Herbst, E., 2005, {\it A\&A}, {\bf 434}, 599
\bibitem{} Charnley, S.B., 2001, {\it ApJ}, {\bf 562L}, 99
\bibitem{} Farebrother, A.J., Meijer, A.J.H.M., Clary, D.C. and Fisher, A.J., 2000, {\it Chem. Phys. Lett.} {\bf 319}, 303
\bibitem{} Gibb, E. L. {\it et al.}, 2000, {\it ApJ}, 536, 347
\bibitem{} Gibb, E. L., Whittet, D. C. B., Boogert, A. C. A. and Tielens, A. G. G. M., 2004,ApJS, 151, 35
\bibitem{} Green, N.J.B {\it et al.} 2001, {\it A\&A}, {\bf 375}, 1111
\bibitem{} Helmich, F. P. {\it et al.} 1996, {\it A\&A}, {\bf 315L}, 173
\bibitem{} Hollenbach, D. and Salpeter, E. E., 1970, {\it J. Chem. Phys.}, 53, 79
\bibitem{} Hollenbach, D., Werner, M.W. and Salpeter, E. E., 1971,  {\it ApJ} {\bf 163}, 165
\bibitem{} Hasegawa, T. and Herbst, E., 1993, {\it MNRAS}, {\bf 261}, 83
\bibitem{} Hasegawa, T., Herbst, E. and Leung, C.M., 1992, {\it APJ}, {\bf 82}, 167
\bibitem{} Katz, N., Furmann, I., Biham, O., Pironello, V. and Vidali, G., 1999, {\it ApJ} {\bf 522}, 305
\bibitem{} Mathis, J.S., Rumpl, W., and  Nordsieck, K.H., 1977, ApJ, 217, 425 
\bibitem{} Pirronello, V., Biham, O., Liu, C., Shena, L. and Vidali, G., 1997a, 
{\it ApJ}, {\bf 483L}, 131
\bibitem{} Pirronello, V., Liu, C., Shena, L. and Vidali, G., 1997b, {\it ApJ}, {\bf 475L}, 69
\bibitem{} Pirronello, V., Liu, C., Riser, J.E. and Vidali, G., 1999, {\it A\&A}, {\bf 344}, 681
\bibitem{} Pontoppidan, K. M., van Dishoeck, E. F. and Dartois, E., 2004, A\&A, 426, 925
\bibitem{} Roberts, H. and Herbst, E., 2002, A\&A, 395, 233
\bibitem{} Snell, R. L. et al., 2000, {\it ApJ},{\bf 539L}, 101
\bibitem{} Stantcheva, T., Caselli, P. and Herbst, E., 2001{\it A\&A}, {\bf 375}, 673
\bibitem{} Stantcheva, T., Shematovich, V. I. and Herbst, E., 2002 {\it A\&A}, {\bf 391}, 1069
\bibitem{} Takahashi, J., Matsuda, K. and Nagaoka, M.,  1999, {\it ApJ}, {\bf 520}, 724
\bibitem{} Tielens, A. G. G. M and Allamandola, L.J., 1987, 
in Interstellar Processes, eds. D.J. Hollenbach and J. H. A. Thronson (Dordrecht), 397
\bibitem{} Tielens, A. G. G. M. and Hagen, W., 1982, {\it A\&A}, {\bf 114}, 245
\bibitem{} Tielens, A. G. G. M., Tokunaga, A.T., Geballe, T.R. and Baas, F., 1982, {\it A\&A}, {\bf 114}, 245
\bibitem{} Tielens, A. G. G. M., Tokunaga, A.T., Geballe, T.R. and Baas, F., 1991, {\it ApJ}, {\bf 381}, 181
\bibitem{} van Dishoeck, E. F. and Helmich, F. P., 1996, {\it A\&A}, {\bf 315l}, 177
\bibitem{} Van der Tak, F. F. S., van Dishoeck, E. F. and Caselli, P., 2000 {\it A\&A}, {\bf 361}, 327
\bibitem{} Watson, W. D. and Salpeter, E. E., 1972 {\it ApJ}, {\bf 174}, 321
\bibitem{} Watson, W. D. and Salpeter, E. E., {\it ApJ}, 1972, 175, 659



\end{thebibliography}
\end{document}